\documentclass[a4paper]{jpconf}
\usepackage{graphicx}
\usepackage[latin1]{inputenc}
\usepackage{amsfonts}
\usepackage{amsxtra}
\usepackage{exscale,amsthm}
\usepackage{latexsym,rotating,fancyhdr}
\usepackage{graphicx,psfrag,subfigure,epsfig,float}
\usepackage{epic,ecltree}
\usepackage{cite}
\usepackage{bbm}
\usepackage{amsmath}
\usepackage{amssymb}
\usepackage{latexsym}
\usepackage{amsfonts}
\usepackage{exscale}
\usepackage{graphicx}
\usepackage{psfrag}
\usepackage{epsfig}
\usepackage{fancybox}
\usepackage{dsfont}
\usepackage{url}
\usepackage[NoDate]{currvita}
\usepackage{ifthen}

\newcommand{\be}{\begin{equation}}
\newcommand{\ee}{\end{equation}}

\newcommand{\bea}{\begin{eqnarray}}
\newcommand{\eea}{\end{eqnarray}}

\newcommand{\R}{{\kern+.25em\sf{R}\kern-.78em\sf{I} \kern+.78em\kern-.25em}}
\newcommand{\RR}{{\kern+.25em\sf{R}\kern-.6em\sf{I} \kern+.6em\kern-.25em}}
\newcommand{\N}{{\kern+.25em\sf{N}\kern-.78em\sf{I} \kern+.78em\kern-.25em}}
\newcommand{\C}{{\kern+.25em\sf{C}\kern-.50em\sf{I} \kern+.50em\kern-.25em}}

\begin{document}
\title{New insight into the Berezinskii-Kosterlitz-Thouless phase transition}

\author{Urs Gerber$^{\rm \, 1}$, Wolfgang Bietenholz$^{\rm \, 1}$ and Fernando G Rej\'{o}n-Barrera$^{\rm \, 2}$}

\address{$^{\rm \, 1}$  Instituto de Ciencias Nucleares, Universidad Nacional Aut\'{o}noma de M\'{e}xico \\
A.P. 70-543, C.P. 04510 Distrito Federal, Mexico}

\address{$^{\rm \, 2}$ Institute for Theoretical Physics, University of Amsterdam \\
Science Park 904, Postbus 94485, 1090 GL Amsterdam, The Netherlands}

\ead{gerber@correo.nucleares.unam.mx, wolbi@nucleares.unam.mx, f.rejon@student.uva.nl}

\begin{abstract}
We investigate the 2d XY model by using the constraint angle action, which belongs to the class of topological lattice actions. These actions violate important features usually demanded for a lattice action, such as the correct classical continuum limit and the applicability of perturbation theory. Nevertheless, they still lead to the same universal quantum continuum limit and show excellent scaling behavior. By using the constraint angle action we gain new insight into the Berezinskii-Kosterlitz-Thouless phase transition of the 2d XY model. This phase transition is of special interest since it is one of the few examples of a phase transition beyond second order. It is of infinite order and therefore an essential phase transition. In particular, we observe an excellent scaling behavior of the helicity modulus, which characterizes this phase transition. We also observe that the mechanism of (un)binding vortex--anti-vortex pairs follows the usual pattern, although free vortices do not require any energy in the 
formulation of the 
2d XY model 
using the constraint angle action.
\end{abstract}

\section{2d XY model and the constraint angle action}
The 2d XY model has fascinated people since the early seventies, in particular because of its description of systems in condensed matter physics, such as superfluid helium films \cite{Kos73} and superconducting films \cite{Bea79}. Further applications include the Coulomb gas model \cite{Fro81}, Josephson junction arrays \cite{Jja, Roj96} and nematic liquid crystals \cite{Par94}. The 2d XY model is of particular conceptual interest because it involves an essential phase transition, i.e. a phase transition of infinite order. Berezinskii \cite{Ber71, Ber72}, as well as Kosterlitz and Thouless \cite{Kos73} realized that the key to the understanding of this phase transition are the vortices and the anti-vortices.

\begin{figure}[htb]
\begin{center}
\includegraphics[width=0.25\textwidth,angle=0]{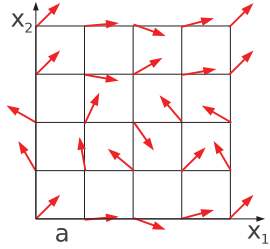}
\caption{Example of a spin configuration on a $4 \times 4$ square lattice with lattice spacing $a$ and periodic boundary conditions.}
\label{exconf}
\end{center}
\end{figure}

In the 2d XY model a classical spin $\vec e_{x}$ is attached to each site $x=(x_1,x_2)$ e.g. on an $L \times L$ square lattice
\be
\vec e_{x} = \left( \begin{array}{c} \cos \phi_{x} \\ \sin \phi_{x}
\end{array} \right) \in \R^{2} \ ,
\ee
such that $| \vec e_{x} | = 1\, , \ \forall x \,$. In Figure \ref{exconf} we show an example of a spin configuration on a $4 \times 4$ square lattice with periodic boundary conditions. We also indicate the lattice spacing $a$.
The standard lattice action reads
\be  \label{stanact}
S [ \vec e \, ] = \sum_{\langle x y \rangle} 
s( \vec e_{x} , \vec e_{y} ) \ , \quad
s( \vec e_{x} , \vec e_{y} ) =
\beta \Big( 1 - \vec e_{x} \cdot \vec e_{y} \Big) 
= \beta \Big( 1 - \cos ( \phi_{x} - \phi_{y} ) \Big)
\ , \ \,
\ee 
with a positive inverse coupling constant $\beta > 0$. The sum runs over all nearest neighbor sites $x, y$. The standard lattice action is obtained by discretizing the derivatives of the continuum formulation of the 2d XY model. Due to the principle of universality, other lattice actions with additional spin couplings, such as the Villain action \cite{Fro81}, lead to the same universal quantum continuum limit. The 2d XY model was also investigated by using the step action, which is not a continuous function in the angle $\phi_{x}$. Instead it is given by a step function
\be 
S [ \vec e \, ] = \sum_{\langle x y \rangle} 
s (\phi_{x}, \phi_{y} ) \ , \quad
s (\phi_{x}, \phi_{y} ) = \left\{ \begin{array}{cccc}
- \beta &&& |\phi_{x} - \phi_{y} | < \pi / 2 \\
\ ~ \beta &&& {\rm otherwise} \end{array} \right. \ .
\ee
It turns out that the step action indeed has the same universal quantum continuum limit as the standard lattice action \cite{KenIrv,  OlsHol}. Here we study the 2d XY model by using the constraint angle action. This action belongs to the class of topological lattice actions. We define topological lattice actions as lattice actions that are invariant under (most) small deformations of a spin configuration \cite{Bie10}. In particular, for the constraint angle action the contribution of a nearest neighbor spin pair amounts to
\be  \label{conact}
s (\vec e_{x}, \vec e_{y} ) = \left\{ \begin{array}{cccc}
0 &&& \vec e_{x} \cdot \vec e_{y} > \cos \delta \\
+ \infty &&& {\rm otherwise} \end{array} \right. \ .
\ee
It does not have any couplings at all. Nearest neighbor spins are constrained to some relative maximum angle $\delta$. All configurations that violate this constraint for at least one spin pair $(\vec e_{x}$, $\vec e_{y})$ have an infinite action (and are therefore excluded), while all other configurations have the same action $S [\vec e \, ] = \sum_{\langle x, y \rangle } s (\vec e_{x}, \vec e_{y} ) = 0$. For $\beta \rightarrow \infty$ the step action corresponds to the constraint angle action with $\delta = \pi / 2$.

It should be pointed out that the idea of using a constraint angle action is not new. Patrascioiu and Seiler \cite{Pat92a,Pat92b} as well as Aizenman \cite{Aiz94} have used an action with an angle constraint to simplify the proof for the existence of a massless phase in the 2d XY model. Furthermore, Patrascioiu and Seiler have also used a constraint angle action in their search for a massless phase in the 2d $O(3)$ model \cite{Pat02}. References \cite{Pat02,Has96} presented numerical studies of the constraint angle action.

In the following sections we discuss our results and the new insight into the Berezinskii-Kosterlitz-Thouless phase transition. This proceeding contribution summarizes our publication \cite{Bie13}. 

\newpage

\section{Berezinskii-Kosterlitz-Thouless phase transition}
\begin{figure}[htb]
\begin{center}
\vspace*{-1cm}
\includegraphics[width=0.5\textwidth,angle=0]{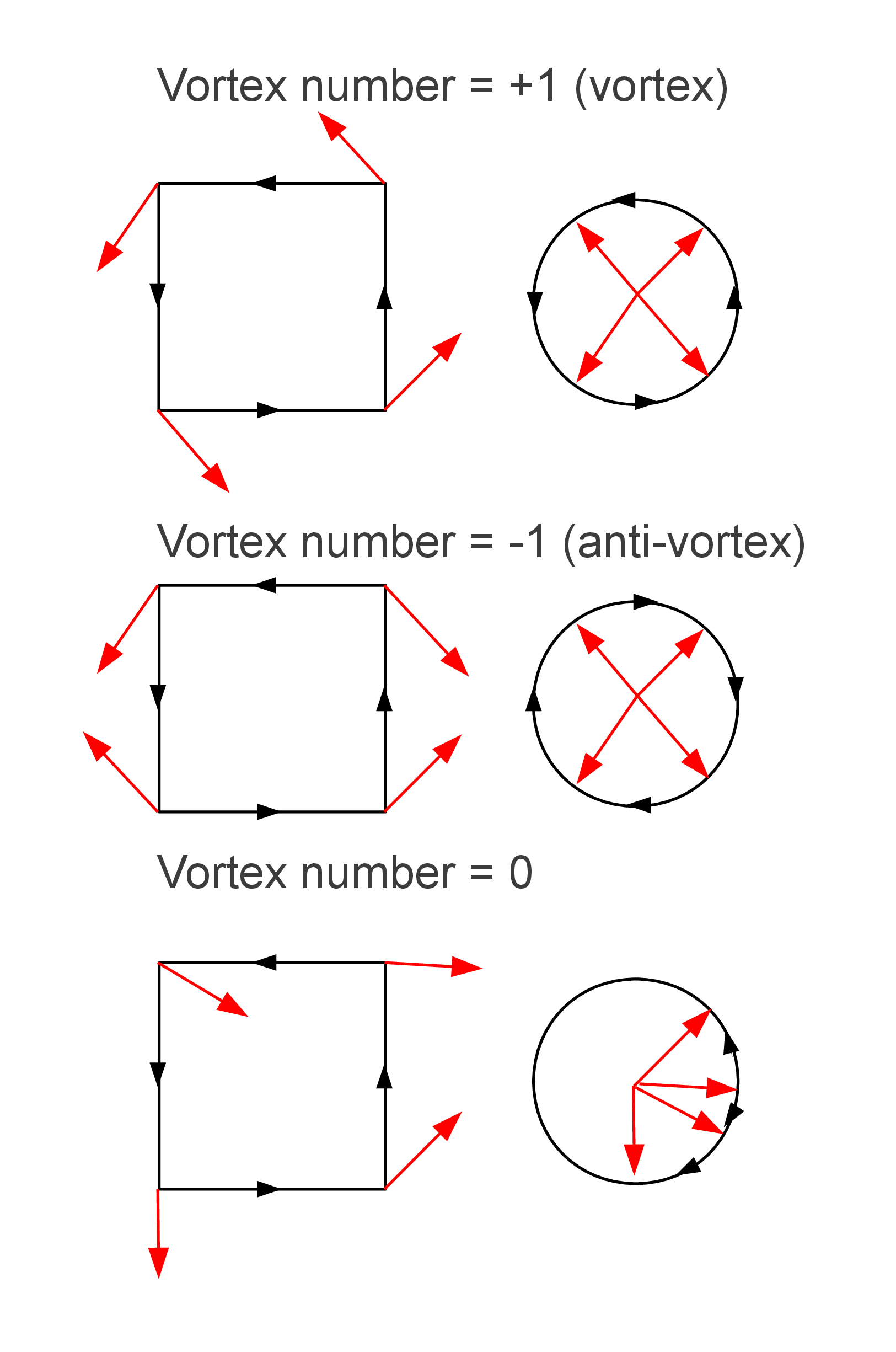}
\caption{Vortices and anti-vortices: On a square lattice each plaquette has a winding number $0$, $+1$ (vortex) or $-1$ (anti-vortex).}
\label{vortices}
\end{center}
\end{figure}

The key to the understanding of the Berezinskii-Kosterlitz-Thouless (BKT) phase transition are the vortices and anti-vortices. On a square lattice, each plaquette has a winding number $0$, $+1$ (vortex) or $-1$ (anti-vortex). This is illustrated in Figure \ref{vortices}.

First an orientation on the plaquettes is chosen. Here we choose it counter-clockwise. The four spins on the plaquette are then mapped on the unit circle. The shortest arc between the neighboring spins are identified. If the circle is closed counter-clockwise we identify a vortex. On the other hand, if the circle is closed clockwise we deal with an anti-vortex. The last possibility is that no circle is covered. In this case we have neither a vortex nor an anti-vortex. Only these three cases are possible. In particular it is not possible that the circle is covered more than once.

\begin{figure}[htb]
\begin{center}
\includegraphics[angle=0,width=0.35\linewidth]{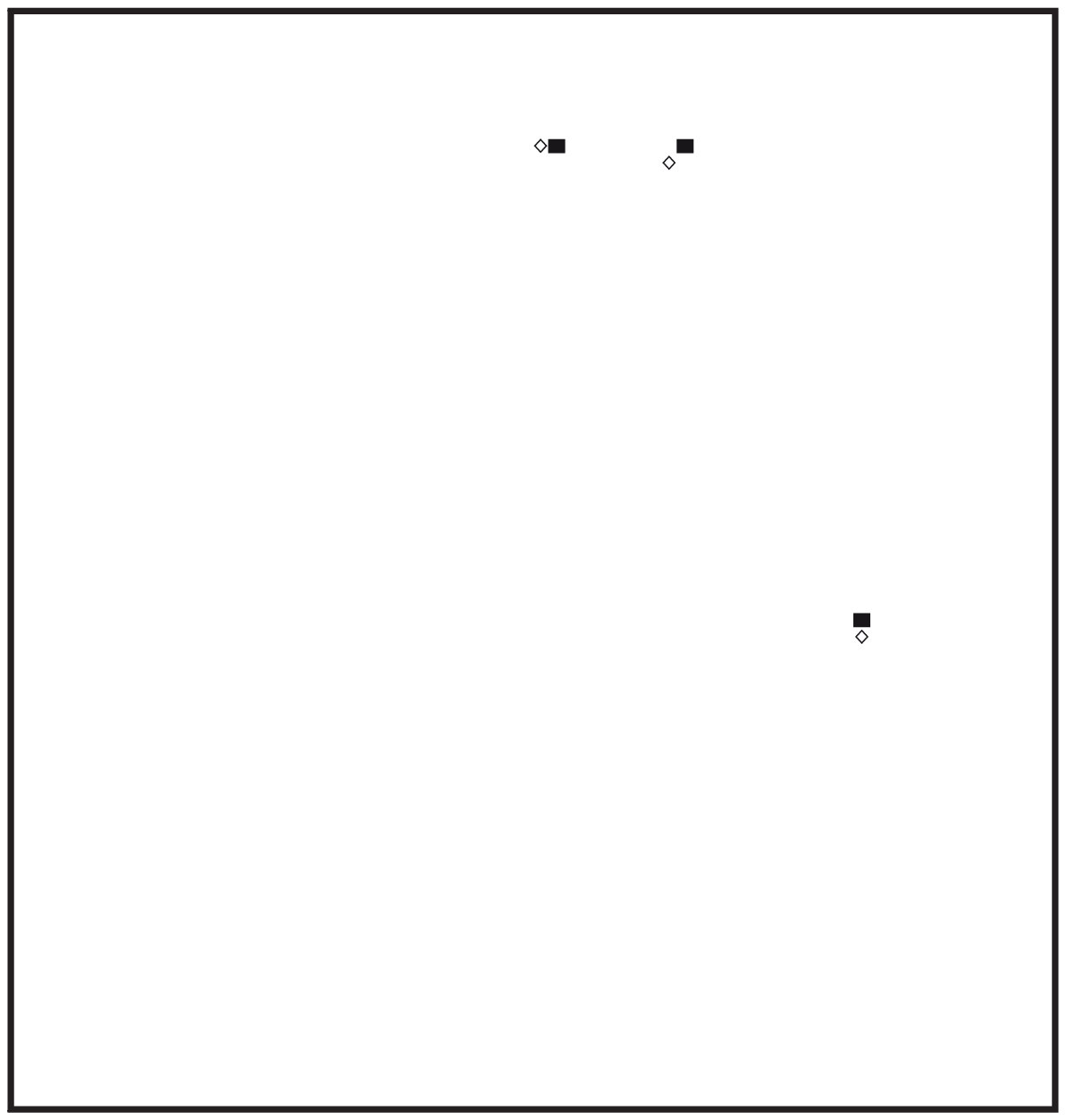}
\hspace*{2mm}
\includegraphics[angle=0,width=0.35\linewidth]{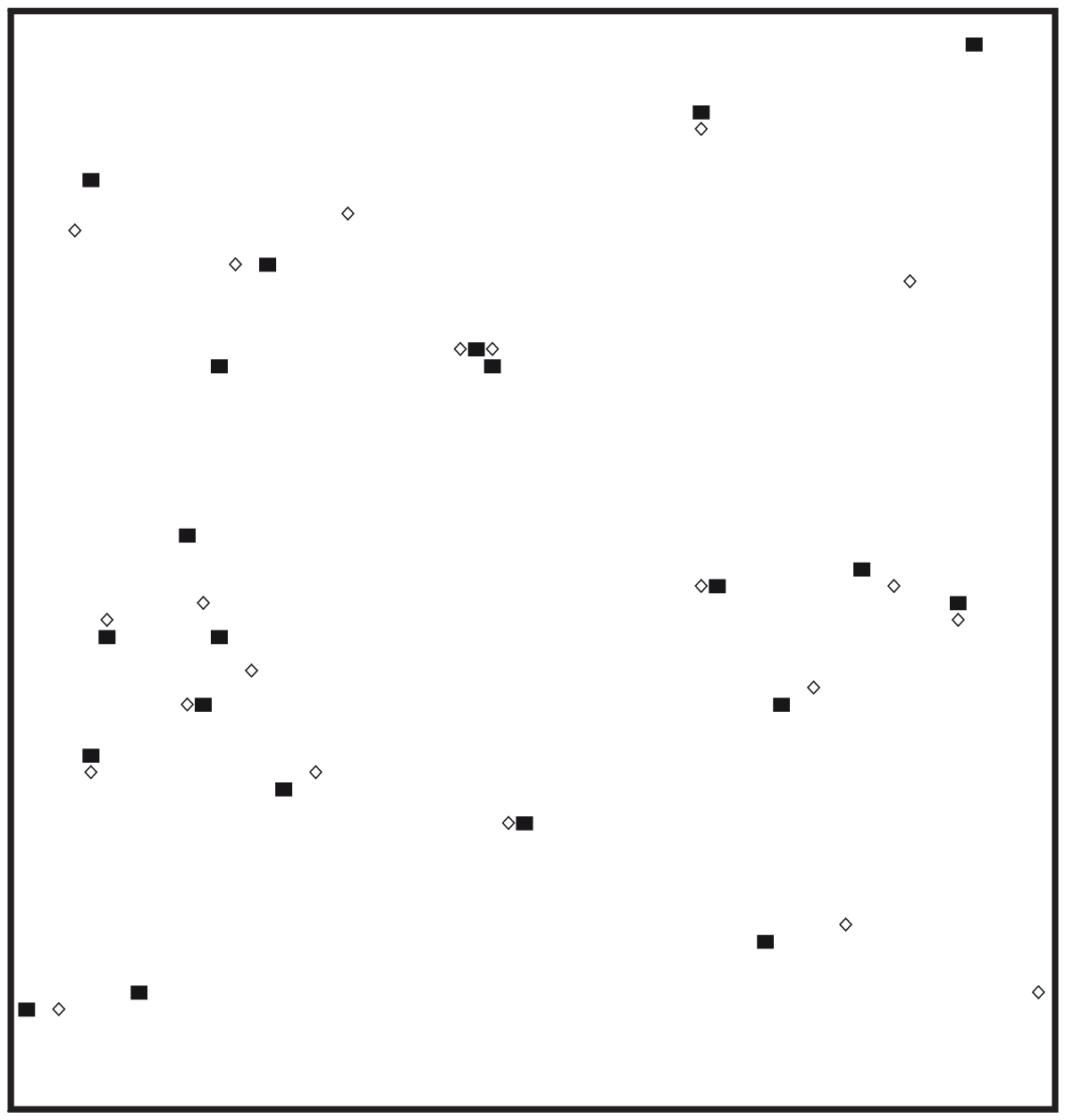}
\vspace*{2mm} \\
\includegraphics[angle=0,width=0.35\linewidth]{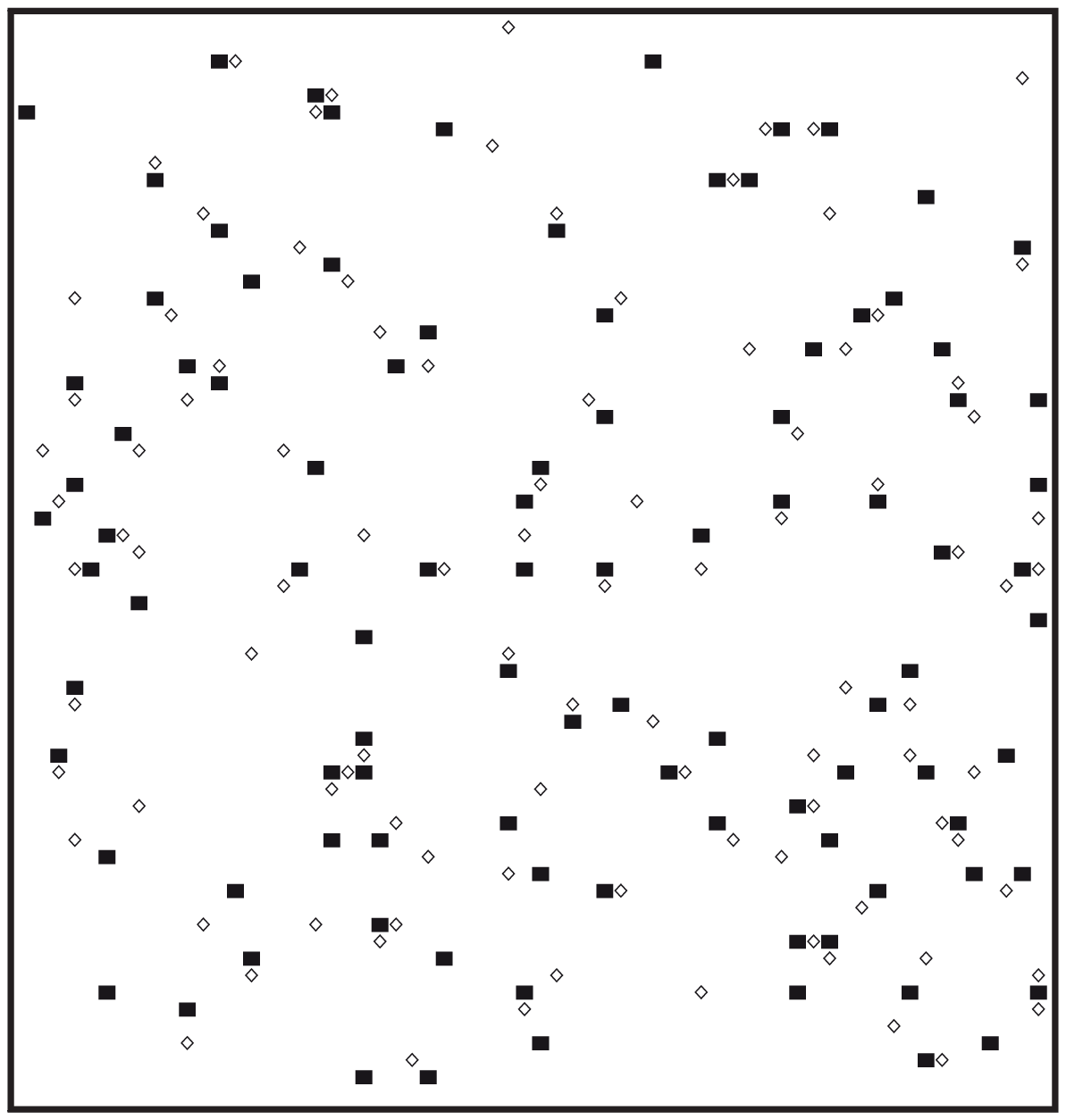}
\hspace*{2mm}
\includegraphics[angle=0,width=0.35\linewidth]{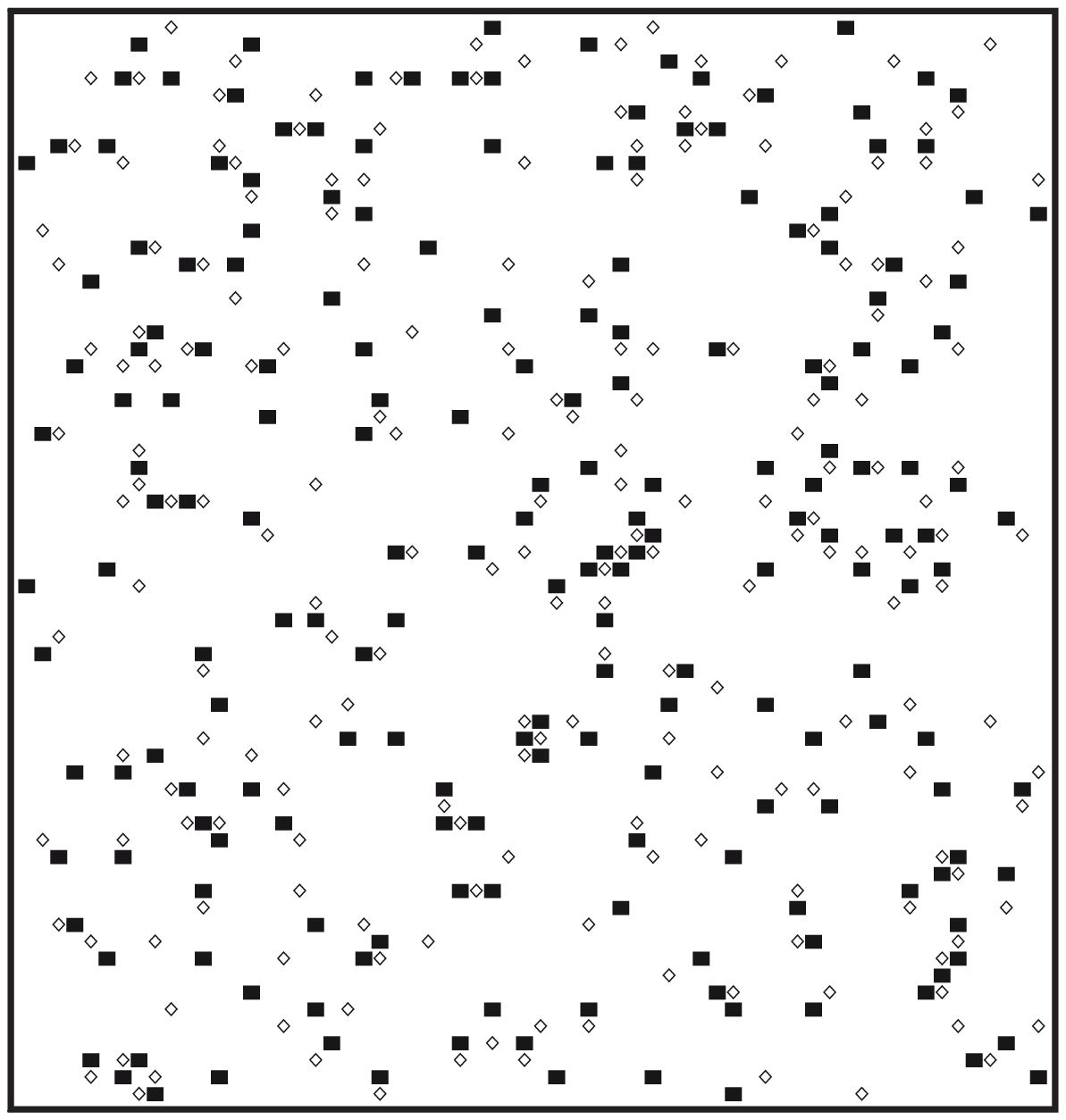}
\caption{Typical configurations of a $L \times L$ lattice with $L=64$ at $\delta =1.85$, $2$ (upper plots), $2.15$ and $2.3$ (lower plots). The vortices are illustrated as black squares and the anti-vortices as white diamonds.}
\label{vorticesconf}
\end{center}
\end{figure}

In Figure \ref{vorticesconf} we show typical configurations of a $L \times L$ lattice with $L=64$ by using the constraint angle action. The vortices are illustrated as black squares whereas the anti-vortices as white diamonds. On top left, for $\delta=1.85$ we observe that most of the vortices and anti-vortices appear as tightly bound vortex--anti-vortex pairs. In fact it is the density of free vortices (not bound in pairs) that is crucial for the BKT phase transition.  There is clearly some ambiguity for an explicit definition of the free vortex density. A possibility is to count those vortices which are not accompanied by any anti-vortex (and vice versa) within some Euclidean distance $r$. In the next section we discuss the free vortex density quantitatively.

For $\delta < \delta_{\rm c}$ the free vortex density is low and a long-range order emerges. The tightly bound vortex--anti-vortex pairs appear neutral from a large-scale perspective. The correlations only decay with a power law, and the correlation length $\xi$ is infinite. Since $m=1/\xi$, this phase is called massless phase. The qualitative situation for the standard lattice action at $\beta > \beta_{\rm c}$ is the same.

If we reach the region where $\delta \gtrsim \delta_{\rm c}$, the vortex--anti-vortex pairs dissociate and the free vortex density jumps up significantly. This destroys the long-range order, the correlation length $\xi$ becomes finite and we enter the massive phase. Again, for the standard lattice action the situation is qualitatively the same for $\beta \lesssim \beta_{\rm c}$.

The infinite order phase transition is characterized by the exponential divergence of the correlation length $\xi$ at $\beta \lesssim \beta_{\rm c}$
\be 
\xi \propto \exp \Big( \frac{\rm const.}{(\beta_{\rm c} 
- \beta )^{\nu}} \Big) \ , \qquad \nu_{\rm c} = 1/2 \ .
\ee
For the constraint angle action at $\delta \gtrsim \delta_{\rm c}$ the divergence of the correlation 
length can be fitted well to \cite{XYtopact}
\be 
\xi \propto \exp \Big( \frac{\rm const.}
{\sqrt{\delta - \delta_{\rm c}}} \Big) \ ,
\ee
which results in a critical angle $\delta_{\rm c} = 1.775(1)$ \cite{Bie13}. For the standard lattice action the BKT phase transition occurs at $\beta_{\rm c} = 1.1199(1)$ \cite{HasPin}.
\newpage
\section{Vortex density and vortex--anti-vortex pairs}
In the previous section we discussed the vortices and the anti-vortices in a qualitative way. In this section we would like to discuss them quantitatively. We perform Monte Carlo simulations of the 2d XY model in the formulation with the constraint angle action. To update the spins we use the very efficient Wolff cluster algorithm \cite{Wol89}.

\begin{figure}[htb]
\begin{center}
\includegraphics[width=0.345\textwidth,angle=270]{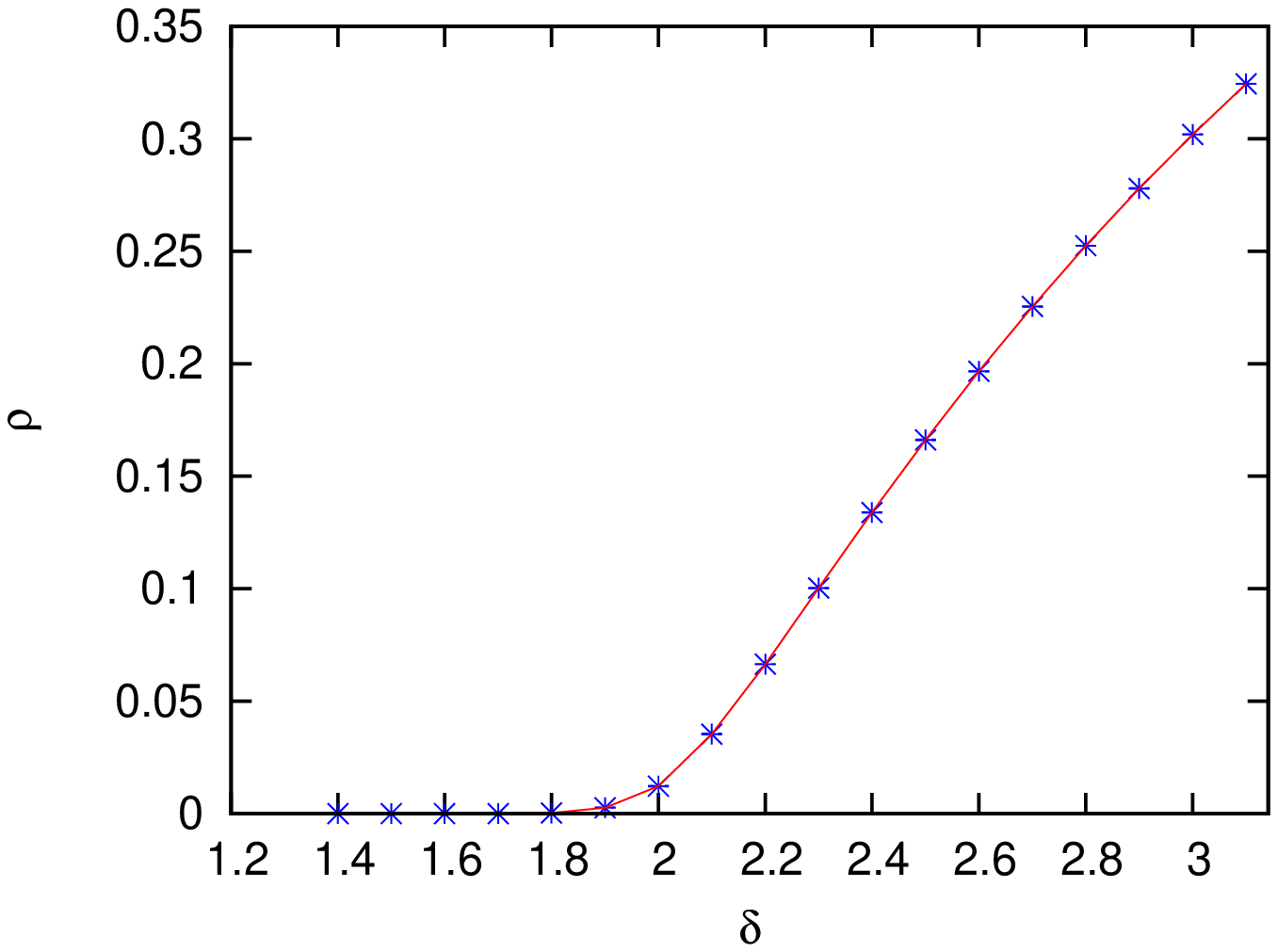}
\includegraphics[width=0.345\textwidth,angle=270]{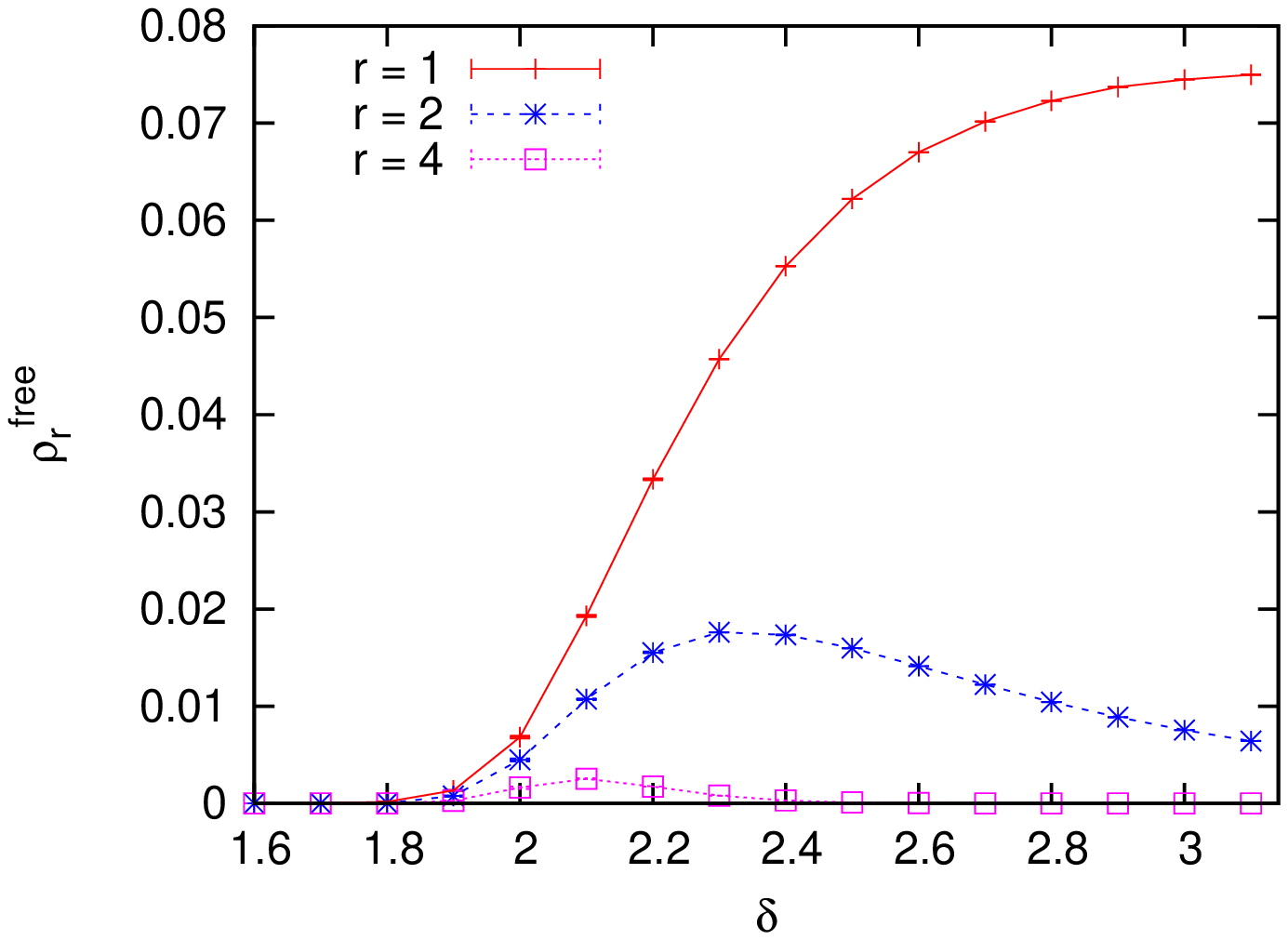}
\caption{Left: Vortex density $\rho$, which is given by the number of plaquettes with $|v_{\Box}| = 1$, divided by the volume. Right: Free vortex density, which can be defined as the density of vortices which are not accompanied by any anti-vortex (and vice versa) within some Euclidean distance $r$. Both plots are obtained on a $L \times L$ lattice with $L=128$ using the constraint angle action.}
\label{vortdens}
\end{center}
\end{figure}

Since we use periodic boundary conditions, Stokes' Theorem implies that the total vorticity always vanishes
\be
\sum_{\Box} v_{\Box} = 0 \ .
\ee 
We measured the vortex density $\rho$, which is given by the number of plaquettes with $|v_{\Box}| = 1$, divided by the volume.
In Figure \ref{vortdens} we show the results on a $L \times L$ lattice with $L=128$ using the constraint angle action.
For $\delta> \pi / 2$ vortices (and anti-vortices) are possible. We mentioned before that it is the free vortex density (rather than the vortex density), which drives the BKT transition. In Figure \ref{vorticesconf} we indeed see qualitatively that for values $\delta \gtrsim \delta_{\rm c}$ the tightly bound vortex--anti-vortex pairs dissociate and the vortices and anti-vortices have no opposite neighbor. We show the free vortex densities for $\rho_{r}^{\rm free}$ for $r=1$, $r=2$ and $r=4$ in Figure \ref{vortdens}. For $r=1$ the curve of the free vortex density increases monotonously for $\delta> \pi / 2$. For $r=2$ and $r=4$ the curves decrease above some value of $\delta$. This can be explained by the fact that the system is filled up by many 
vortices and anti-vortices and therefore the probability to have an opposite partner within the radius $r$ rises again. For $\delta \gtrsim \delta_{\rm c}$ we observe that all the curves grow rapidly which confirms the unbinding of the vortex--anti-vortex pairs. 

Furthermore, we studied the pair formation of the vortices and anti-vortices in the following way. Given a configuration, we first identify its $N$ vortices and $N$ anti-vortices, and search for the optimal pairing. This optimization minimizes the quantity
\be \label{D2}
D^{2} = \frac{1}{N} \, \sum_{i=1}^{N} d_{{\rm VA},\, i}^{\, 2} \ \, ,
\ee
where $d_{{\rm VA},\, i}$ are the Euclidean distances that separate the vortex--anti-vortex partners. The direct method of checking all possibilities is only applicable up to $N \approx 14$. We work again at lattice size $L = 128$ with constraint angles up to $\delta =2.05$, where typically $N$ is close to $200$. In order to still identify the optimal pairing (with high probability), we applied the technique of simulated annealing \cite{anneal}.
As a further reference quantity, we add the quantities
\be \
R_{\rm VV}^{2} = \frac{2}{N} \, \sum_{i=1}^{N/2} d_{{\rm VV},\, i}^{\, 2} \quad 
{\rm and} \quad
R_{\rm AA}^{2} = \frac{2}{N} \, \sum_{i=1}^{N/2} d_{{\rm AA},\, i}^{\, 2} \ ,
\ee
where $d_{{\rm VV},\, i}$ ($d_{{\rm AA},\, i}$) are the distances between two vortices (two anti-vortices)
We show the ratio between the quantities $D^{2}$, $R_{\rm VV}^{2}$ and  $R_{\rm AA}^{2}$ for simulated configurations, and for random 
distributions, with $N$ vortices and $N$ anti-vortices.
\begin{figure}[htb]
\begin{center}
\includegraphics[width=0.55\textwidth,angle=270]{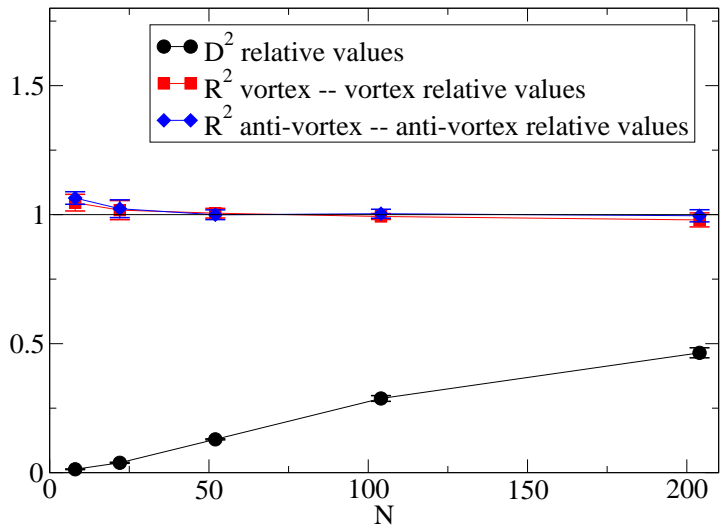}
\caption{Ratio between the quantities $D^{2}$, $R_{\rm VV}^{2}$ and  $R_{\rm AA}^{2}$ for simulated configurations, and for random 
distributions, with $N$ vortices and $N$ anti-vortices on a $L \times L$ lattice with $L=128$ using the constraint angle action.}
\end{center}
\end{figure}
We observe a strong trend towards pair formation at small $N$. For increasing $N$ this trend fades away.

The established picture of the BKT transition also implies a sizable vorticity anti-correlation over short distances in the massless
phase, in particular over distance $1$. We consider the vorticity correlation function
\be \label{qcorrfunc}
C(r) = \langle \, v_{\Box , (x_{1},x_{2})} \, v_{\Box , (x_{1}+r,x_{2})} \, 
\rangle |_{|v_{\Box , (x_{1},x_{2})}|=1}
\ee
over distances $r = 1$, 2 and 3, at a set of constraint angles
$\delta \geq \delta_{\rm c}$.
\begin{figure}[htb]
\begin{center}
\includegraphics[width=0.55\textwidth,angle=270]{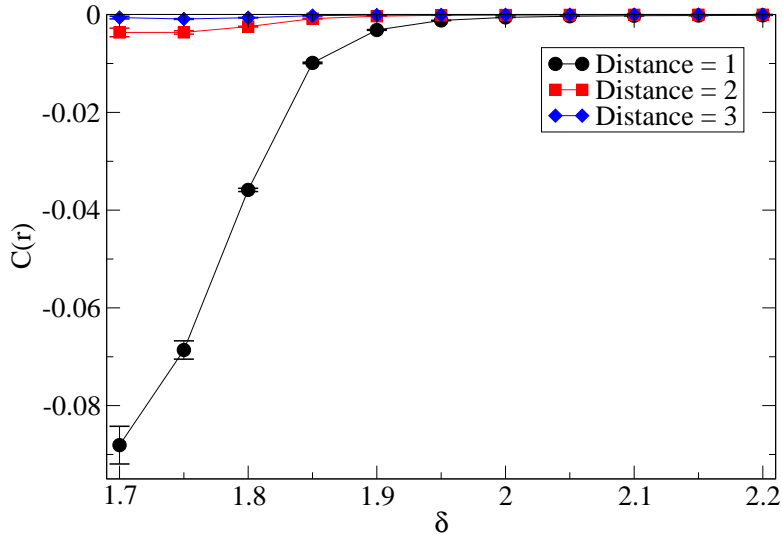}
\caption{Vorticity correlation function $C(r)$ in eq.\ (\ref{qcorrfunc}) over distances $r=1, 2$ and $3$ as a function of the constraint angle $\delta$. We observe a strong anti-correlation over the smallest distance $r=1$ at $\delta \lesssim \delta_c$.  This is consistent with the picture of the formation of vortex---anti-vortex pairs. As $\delta$ increases the pairs tend to dissociate.}
\end{center}
\end{figure}
Indeed, we confirm a strong anti-correlation at distance 1 around $\delta_c$, which decreases rapidly as $\delta$ increases. The negative values for $C(r)$ can be explained since neighboring plaquettes contribute with opposite sign to the vorticity correlation function.

\newpage
\section{Helicity modulus}
The helicity modulus is a well-known quantity in condensed matter physics. It is proportional to the superfluid density and in the literature sometime also called 'spin stiffness' or 'spin rigidity'. It is a measure for the sensitivity of a system to torsion, in particular to a variation of a twist in the boundary conditions. Furthermore, besides the critical exponents, this
is often a useful indicator to characterize a universality class.

For a non-topological lattice action like the standard lattice action, the helicity modulus $\Upsilon$ can be defined as
\be
\Upsilon 
= \frac{\partial^{2}}{\partial \alpha^{2}} 
F (\alpha ) |_{\alpha = 0} \ ,
\ee
where $F = - \frac{1}{\beta} \ln Z$ is the free energy, and $Z$ the partition function. The boundary conditions are assumed to be periodic in one direction and twisted with the angle $\alpha$ in the other one. The helicity modulus corresponds to the curvature in the minimum at $\alpha =0$.

Regarding topological lattice actions, and in particular for the constraint angle action this definition is problematic. The constraint angle action has no couplings and therefore the above definition is inappropriate. Hence we consider a modification and define the dimensionless helicity modulus as
\be
\bar \Upsilon :=  \beta \, \Upsilon \ .
\ee
The critical value of the helicity modulus for the 2d XY model in a square volume was first predicted analytically in \cite{Nel77, Min81} to be
$\bar \Upsilon_{\rm c,\, theory} = \frac{2}{\pi} . $
Later it was noticed in \cite{Pro00} (see also \cite{Has05}) that a tiny correction (below $0.2$ per mille) due to winding 
configurations has to be taken into account,
\be \label{BKTtheory}
\bar \Upsilon_{\rm c,\, theory} = 
\frac{2}{\pi} \Big( 1 - 16 \pi e^{-4 \pi} \Big)
\simeq 0.636508 \ .
\ee

Using the standard lattice action, $\bar \Upsilon$ at $\alpha =0$ can be evaluated in a convenient way such that the generation of configurations can be restricted to periodic boundary conditions. In this way the most extensive study has been worked out by Hasenbusch in \cite{Has05}. He performed simulations on lattices up to $L=2048$, where he obtained at $\beta_{\rm c} = 
1.1199$ the value $\bar \Upsilon_{\rm c} = 0.67246(10)$. Compared to the analytic prediction this is $5.6 ~\%$ too large. (For convenience, we denote the dimensionless helicity modulus at the critical parameter by $\bar \Upsilon_{\rm c}$, even in finite volume.) In order to obtain an infinite volume extrapolation, he fitted his results for various sizes $L$ to the form
\be
\bar \Upsilon_{\rm c} (L) = \bar \Upsilon_{\rm c,\, theory}
+ \frac{c_{1}}{\ln L + c_{2}} \ ,
\ee
with free parameters $c_{1}, \ c_{2}$, which worked decently. Theoretical predictions in the spin wave limit \cite{Has05} and based on the renormalization group \cite{Pel13} argue for the universality of $c_{1}$ and predict the value $c_{1} \simeq 0.3189$. However, it should be mentioned that in this calculation the parameter $\beta$ is inverted. Since the constraint angle action does not involve any $\beta$ parameter, it is not obvious if the universality arguments for $c_{1}$ still apply in this case.

For the constraint angle action the determination of $\bar \Upsilon$ at $\alpha =0$ fails. A small change in $\alpha$ does (in general) not affect $F$ at all (in a finite volume). On the first sight this seems to be problematic. However, $\bar \Upsilon$ can also be evaluated from a histogram for the $\alpha$ values, which describes their probability $p( \alpha )$. In this case $\alpha$ has to be treated as a dynamical variable in the simulation. Its probability density is related to $\bar \Upsilon$ as \cite{OlsHol}
\be 
\bar \Upsilon = - \frac{\partial^{2}}{\partial \alpha^{2}} \ 
\ln p (\alpha) \vert_{\alpha = 0} \ .
\ee
In practice, the idea is to determine the curvature in the maximum of $\ln p (\alpha)$ from a histogram up to moderate $|\alpha | \, $.

For the step action the convenient way of evaluating $\bar \Upsilon$ at $\alpha =0$ is not applicable either. In this case the corresponding histograms for $p(\alpha )$ have been studied in \cite{OlsHol}. The BKT transition is observed around $\beta_{\rm c} \approx 1.2 \dots 1.3$ \cite{KenIrv,OlsHol}. Using twisted boundary conditions, Olsson and Holme measured  $\bar \Upsilon_{\rm c} = 0.663(6)$, at $\beta_{\rm c}$ on a $L \times L$ lattice with $L=256$ lattice. Compared to the standard lattice action this is closer to the BKT value, but still $4.2 \ \%$ too large.

For the constraint angle action we performed Monte Carlo simulations by using the Metropolis algorithm to update the spins and the angle $\alpha$ at the twisted boundary. We also formulated a cluster algorithm \cite{Bie13}. However it turned out that it is not consistent without performing additional accept/reject decisions.

\begin{figure}[htb]
\begin{center}
\includegraphics[width=0.55\textwidth,angle=270]{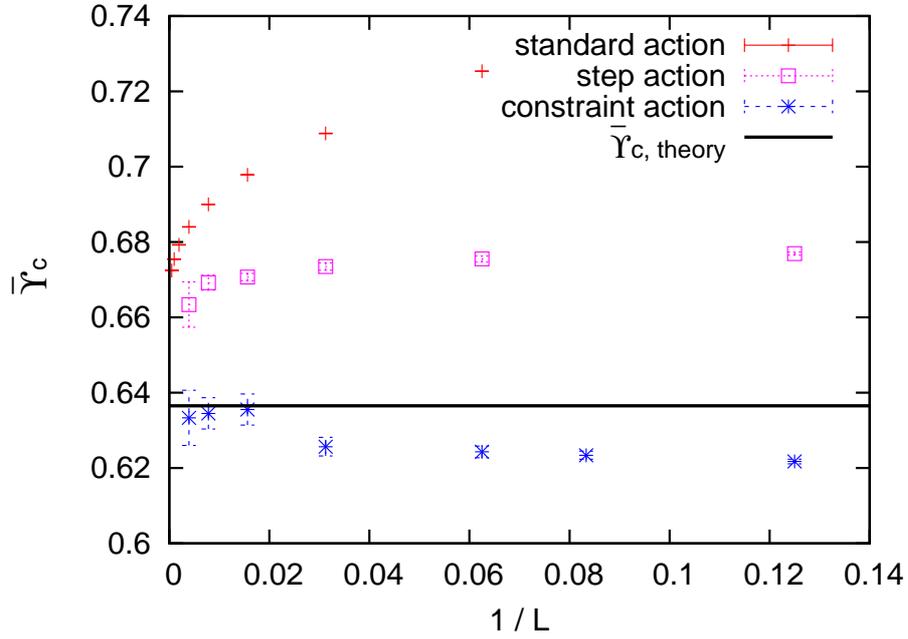}
\caption{Numerical results for $\bar \Upsilon_{\rm c}$ on $L \times L$ lattices at the critical parameter, for the standard lattice action (data from \cite{Has05}), the step action (data from  \cite{OlsHol}) and for the constraint angle action \cite{Bie13}. For the standard lattice action and the step action the results differ significantly from the theoretical BKT value $\bar \Upsilon_{\rm c,\, theory }$ in eq.\ (\ref{BKTtheory}), which could only be attained with extended extrapolations. Only for the constraint action the results agree with the BKT prediction for $L\geq64$, and the deviation is just 1.9\% (2.3\%) even at $L=16$ $(L=8)$.  }
\label{upsilonallactions}
\end{center}
\end{figure}

Using the constraint angle action we measured $\bar \Upsilon_{\rm c}$ in various volumes, focusing on the critical constraint angle $\delta_{\rm c} = 1.775(1)$. We simulated the model on $L \times L$ lattices in the range
$L = 8 \dots 256$ with dynamical boundary conditions. Our results are illustrated in Figure \ref{upsilonallactions}, and compared with those for the standard lattice action \cite{Has05} and for the step action \cite{OlsHol}.

The deviation of $\bar \Upsilon_{\rm c}$ from the theoretical value $\bar \Upsilon_{\rm c,\, theory }$ is less than $ 2.4 \ \%$ for all sizes $L$ that we considered. Starting from $L \geq 64$ our results confirm the prediction within the errors.
In view of earlier attempts to measure $\bar \Upsilon_{\rm c}$ with other lattice actions, this observation is highly remarkable.

\begin{figure}[htb]
\begin{center}
\includegraphics[width=0.6\textwidth,angle=0]{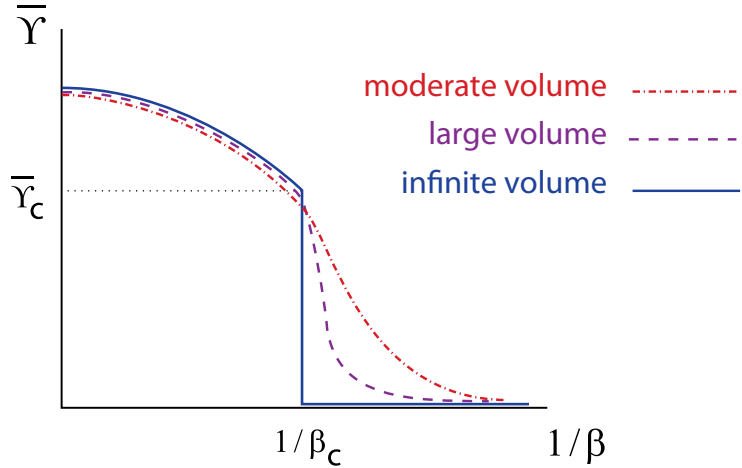}
\caption{Qualitative picture of the expected coupling dependence of the helicity modulus in different volumes. In finite volume the function $\bar \Upsilon (\beta )$ is continuous, but for increasing size $L$ the jump to $0$ is approximated better and better.}
\label{upsilonquali}
\end{center}
\end{figure}

Let us now have a look at the so-called helicity gap. For an infinite volume, the BKT theory predicts a discontinuity
of the helicity modulus. As soon as the coupling exceeds its critical value, $\bar \Upsilon$ drops to $0$. In finite volume the function $\bar \Upsilon (\beta )$ is continuous, but for increasing size $L$ the jump to $0$ is approximated better and better. The situation is illustrated qualitatively for the standard lattice action in Figure \ref{upsilonquali}.

\begin{figure}[htb]
\begin{center}
\includegraphics[width=0.6\textwidth,angle=270]{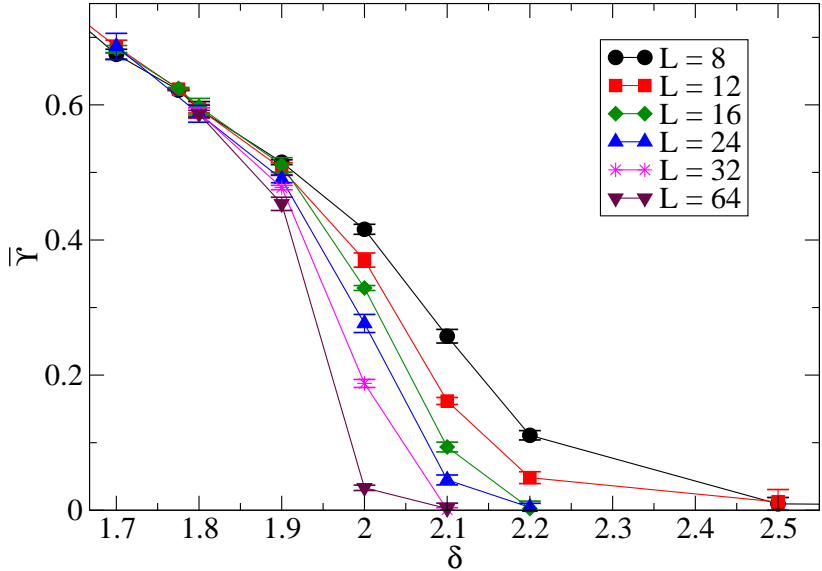}
\caption{Results for $\bar \Upsilon$ in six volumes, over a range of $\delta$ angles, which includes $\delta_{\rm c} \simeq 1.775$. We observe the expected trend towards a jump down to $0$ above 
$\delta_{\rm c}$, in analogy to the schematic Figure \ref{upsilonquali}.}
\label{upsilonquanti}
\end{center}
\end{figure}

For the standard lattice action \cite{Min03} and for the step action \cite{OlsHol} the observations are compatible with this property. We expect the same behavior for the constraint angle action, where $\bar \Upsilon (\delta )$ should jump to $0$ in an infinite volume when $\delta$ exceeds $\delta_{\rm c}$. As a test, we measured
$\bar \Upsilon (\delta )$ in various volumes. We show the results in Figure \ref{upsilonquanti}. Indeed, the data for the constraint angle action are also compatible with the expected qualitative picture.

\newpage
\section{Conclusions}
Using the formulation of the constraint angle action, we have investigated the Berezinskii-Kosterlitz-Thouless phase transition of the 2d XY model. To this end, we performed Monte Carlo simulations using dynamical boundary conditions. We confirmed the value of the dimensionless helicity modulus $\bar \Upsilon_{\rm c,\, theory}$, which was predicted at a BKT phase transition. Using the constraint angle action, the finite size effects are modest, in contrast to other lattice actions such as the standard lattice action and the step action. In particular, the value of $\bar \Upsilon_{\rm c}$ remains close to the BKT value down to volumes as small as $16 \times 16$, where the deviation is just $1.9\%$. These facts eliminate any doubt that the constraint angle action belongs to the same universality class as the conventional lattice actions, which involve spin couplings (such as the standard lattice action).

Furthermore, we verified the picture of vortex--anti-vortex pair (un)binding as the mechanism behind the BKT transition. Our results for the density of free vortices and anti-vortices (without an opposite partner up to some distance), for the vorticity correlation, as well as for the sum over pair separations squared, are all compatible with this picture.

It should be pointed out that the validity of this mechanism is highly non-trivial. Free vortices do not cost any energy (if the constraint allows them). Their suppression in the range $\pi/2 < \delta \lesssim \delta_{\rm c}$ can only be explained by the combinatorial frequency of configurations carrying different vorticities. In this view, our results deviate from the established picture, since they demonstrate that a BKT transition can occur even without any Boltzmann factor suppression of free vortices.

\newpage

\noindent
\ack

We thank Michael B\"{o}gli who has contributed to this work at an early stage. We further thank Uwe-Jens Wiese for instructive discussions, and Martin Hasenbusch, Silvano Romano, Erhard Seiler and Ulli Wolff for 
interesting remarks. 

This work was supported by the Mexican {\it Consejo Nacional de Ciencia 
y Tecnolog\'{\i}a} (CONACyT) through project 155905/10 ``F\'{\i}sica 
de Part\'{\i}culas por medio de Simulaciones Num\'{e}ricas'' and 
through the scholarship 312631 for graduate studies, as well as 
DGAPA-UNAM. The simulations were performed on the cluster of the 
Instituto de Ciencias Nucleares, UNAM. We would like to thank Luciano 
D\'{\i}az and Enrique Palacios for technical support.

\end{document}